\documentstyle{article}
\begin{document}
\unitlength1.0cm
\begin{titlepage}
 \hfill MZ-TH/92-50, hep-ph/9212254\\[1cm]
 \begin{center}
  {\Large {\bf 2-Loop Integrals in The Standard Model
  }}\\[2cm]
  {\large
  D. Kreimer\footnote{
  e-mail: kreimer@vipmza.physik.uni-mainz.de}}\\[1cm]
  Institut f\"ur Physik, Johannes Gutenberg Universit\"at,\\
  Staudinger Weg 7, Postfach 3980, D-6500 Mainz, Germany\\[5cm]
 \end{center}
 We introduce in typical examples new methods for the calculation of
 massive loop integrals appearing in the radiative correction calculations
 of the Standard Model.
 \\[1cm]
\begin{center}
 October 1992\\[2cm]
\end{center}
Talk given at the workshop on HEP and QFT, Sochi, Russia, 1992
\end{titlepage}
\section{Introduction}
In the following we will be concerned with the calculation of massive
loop integrals. We will present methods to calculate such integrals
for the 1- and 2-loop case. For the 1-loop case these methods allow for the
calculation of arbitrary rank tensor integrals for arbitrary
$n$-point functions with arbitrary integer powers of the propagators.
The methods for the 2-loop case cover the scalar planar and nonplanar
integrals. The tensor cases can be handled by the usual reduction
to 1-loop integrals or, in a much more direct approach, by integrating
the 'characteristic integrand' of the graph under consideration \cite{new}.

In this article we give only typical introductory examples of these
methods. We refer the reader to the literature to study the methods in
detail \cite{I,II,III,2l2p,2l3p}.

\section{1-loop Integrals}
 \subsection{The Scalar Function}
In this section we calculate the two-point function
\begin{equation}
   I(q^{2})=\int d^{D}l\frac{1}{P_{1}P_{2}},\label{eq1}
\end{equation}
where
\begin{eqnarray*}
   P_{1} & = & l^{2}-m_{1}^{2}+i\varrho\\
   P_{2} & = & (l+q)^{2}-m_{2}^{2}+i\varrho.
\end{eqnarray*}
Here we do not specify the masses $m_{1},m_{2}$, so we want a result which
is valid for the massive as well as for the massless case. We work within
dimensional regularization in the sense of \cite{Coll}, $D=4-2\varepsilon$ is
the complex dimension.

There is only one scalar-product involved in the integrand Eq.(\ref{eq1})
 and so if we
choose to be in the rest frame of $q=q_{0}$, we can separate the
$l_{0}$-integration. No non-trivial angular integration is then involved
in the $(D-1)$-dimensional space-like integration. So we can do $D-2$
angular integrations trivially, leaving the $l_{0}$ and $l_{\bot}$ integrations
to be done. We have denoted the modulus of the space-like part of $l$
by $l_{\bot}$.

The integral Eq.(\ref{eq1}) then becomes
\begin{eqnarray}
   I(q^{2}) & = & 2\frac{\pi^{\frac{D-1}{2}}}{{\rm \Gamma}(\frac{D-1}{2})}
                  \int_{-\infty}^{\infty}dl_{0}\int_{0}^{\infty}dl_{\bot}
                  l_{\bot}^{D-2}\frac{1}{{\cal P}_{1}{
                  \cal P}_{2}}\label{eq2}\\
   \mbox{where} & & {\cal P}_{1}=l_{0}^{2}-l_{\bot}^{2}-m_{1}^{2}
                     +i\varrho\nonumber\\
                & & {\cal P}_{2}=(l_{0}+q_{0})^{2}-l_{\bot}^{2}-m_{2}^{2}
                     +i\varrho.\nonumber
\end{eqnarray}

The $l_{\bot}$-integration in Eq.(\ref{eq2})
can now easily be performed.
The result is
\begin{eqnarray}
   I(q^{2}) & = & C\int_{-\infty}^{\infty}
   dl_{0}\frac{1}{w_{1}^{2}-w_{2}^{2}}
                  \left[
                     w_{1}^{1-2\varepsilon}-w_{2}^{1-2\varepsilon}
                  \right]\nonumber\\
   \mbox{where}\;\; w_{1} & = & \sqrt{-l_{0}^{2}+m_{1}^{2}-i
\varrho}\nonumber\\
                w_{2} & = & \sqrt{-(l_{0}+q)^{2}+m_{2}^{2}-i
\varrho}\nonumber\\
                C & = & \frac{\pi^{\frac{D-1}{2}}}{1/2-\varepsilon}
                        \Gamma(1/2+\varepsilon).
                \label{eq3}
\end{eqnarray}

Now the remaining $l_{0}$-integration becomes
\begin{eqnarray}
 & & I(q^{2}):=\frac{Ce^{(-i\pi(1/2
    -\varepsilon))}}{2q}\int_{-\infty}^{\infty}dl_{0}\left[
                                   \frac{(l_{0}^{2}-m_{1}^{2}+
    i\varrho)^{\frac{1}{2}-\varepsilon}}{q/2+M_{d}+l_{0}}
                                  +\frac{(l_{0}^{2}-m_{2}^{2}+
    i\varrho)^{\frac{1}{2}-\varepsilon}}{q/2-M_{d}-l_{0}}
                                \right] \nonumber\\
 & & M_{d}=\frac{m_{1}^{2}-m_{2}^{2}}{2q},\mbox{\hfill}\label{eq4}
\end{eqnarray}
where we used translation invariance for the second term in Eq.(\ref{eq3}),
translation invariance being
guaranteed by the very definition of dimensional regularization
(${\rm Re}(\varepsilon)$ big enough in intermediate steps). The integral
Eq.(\ref{eq4}) can
be determined by various methods. We will do it in the following by introducing
${\cal R}$-functions \cite{Carl,I,II} which are the typical ingredient of
our 1-loop method.

We modify the integral Eq.(\ref{eq4}) as follows.
\begin{eqnarray}
 I=\frac{Ci\exp(i\pi\varepsilon)
     }{2q}\int_{0}^{\infty}\frac{ds}{\sqrt{s}}\left[
                                  (q/2+M_{d}) \frac{(s-m_{1}^{2}+
   i\varrho)^{\frac{1}{2}-2\varepsilon}}{-(q/2+M_{d})^{2}+s}\right.
   \nonumber\\
   \left.                               +(q/2-M_{d})\frac{(s-m_{2}^{2}+
   i\varrho)^{\frac{1}{2}-2\varepsilon}}{-(q/2-M_{d})^{2}+s}
                                \right],\label{pv}
\end{eqnarray}
which gives in terms of ${\cal R}$-functions
\begin{eqnarray}
   I & = &\frac{i\exp(i\pi\varepsilon)\;C}{2q}
         \left[ (q/2+M_{d})
           {\rm B}(\frac{1}{2},\varepsilon)\right.\nonumber\\
     & &
        \left.
           {\cal R}_{-\varepsilon}(-\frac{1}{2}+\varepsilon,1;-m_{1}^{2}
           +i\varrho,-(q/2+M_{d})^{2})+
          (q/2-M_{d})\times\right. \nonumber\\
     &  & \left. \times {\rm B}(\frac{1}{2},\varepsilon)
           {\cal R}_{-\varepsilon}(-\frac{1}{2}+\varepsilon,1;-m_{2}^{2}
           +i\varrho,-(q/2-M_{d})^{2})\right].\label{eq5}
\end{eqnarray}

The Beta-function in Eq.(\ref{eq5}) involves a pole in $\varepsilon$.
Expanding the
following ${\cal R}$-functions to ${\cal O}(\varepsilon)$
(see \cite{I}) we can express our
 result in terms
of logarithms.
The result is
\begin{eqnarray*}
 I(q^{2}) & = & i\pi^{2}
                \left\{ \frac{1}{\varepsilon}-\gamma-\log\pi+2
                +\frac{1}{2}(1+m_{1}^{2}/q^{2}-
                 m_{2}^{2}/{q^2})\right. \times\\
          & & \times \left[ z_{1}\log\frac{
                         \stackrel{+}{-}(1-z_{1})}{1+z_{1}}-
                            \log m_{1}^{2} \right] \\
          & & +\frac{1}{2}(1-m_{1}^{2}/q^{2}+m_{2}^{2}/q^{2})\times\\
          & & \times \left[ z_{2} \log\frac{
                          \stackrel{+}{-}(1-z_{2})}{1+z_{2}}-
                       \left.     \log m_{2}^{2}\right] \right\},\\
 \mbox{where} & & z_{1}=\sqrt{1-\frac{4m_{1}^{2}q^{2}}{(
                                q^{2}+m_{1}^{2}-m_{2}^{2})^{2}}}\\
              & & z_{2}=\sqrt{1-\frac{4m_{2}^{2}q^{2}}{(
                                q^{2}-m_{1}^{2}+m_{2}^{2})^{2}}}.
\end{eqnarray*}
where the upper (lower) sign has to be used if the argument is in the
right (left) complex half-plane corresponding to Carlson's conventions on
the scaling law for ${\cal R}$-functions and quadratic transformations on
them \cite{Carl}.
This is the well-known result which one also obtains by conventional
methods.

\subsection{The Tensor Functions}
As it is shown in \cite{I} the above reduction allows also to calculate
arbitrary tensor integrals. Here we give only the result.
\begin{eqnarray*}
 I_{jk}(q^{2}) & = & \tilde{C}\;(\; {\rm B}(\varepsilon-k/2-[j]/2,1/2+
                            [j]/2)(q/2+M_{d})^{j+1-[j]}\\
               & &\times (-1)^{j-[j]}
                     {\cal R}_{-\varepsilon+k/2+[j]/2}(
                     -1/2-k/2+\varepsilon,1;z_{1},y_{1})\\
               & + & \sum_{i=0}^{j}{j \choose i}
               (-q)^{j-i}{\rm B}(\varepsilon-k/2-[i]/2,1/2+[i]/2)
               (q/2-M_{d})^{i+1-[i]}\\
               & & \times
               {\cal R}_{-\varepsilon+k/2+[i]/2}
               (-1/2-k/2+\varepsilon,1;z_{2},y_{2})\;)\\
\mbox{where} & & [m]:= \left\{
                              {{m,m\;\;\mbox{even}} \atop
                               {m+1,m\;\;\mbox{odd}}}
                      \right\} \\
             & & \tilde{C} := \frac{\pi^{\frac{D-1}{2}}}{1/2+k/2-
                       \varepsilon}\frac{i\exp(i\pi\varepsilon)\;
                        (-1)^{k/2}}{2q}\times\\
             & & \times\frac{
                       \Gamma(3/2-\varepsilon+k/2)
                       \Gamma(1/2-k/2+\varepsilon)}{
                       \Gamma(3/2-\varepsilon)}\\
             & & z_{1}:=-m_{1}^{2}+i\varrho,y_{1}=-(q/2+M_{d})^{2}\\
             & & z_{2}:=-m_{2}^{2}+i\varrho,y_{2}=-(q/2-M_{d})^{2}.
\end{eqnarray*}
In the above formula for $I_{jk}(q^{2})$ $j$ denotes the number of indices
in the parallel space and $2k$ the number of indices in orthogonal space.
One then has for a tensor integral
\begin{eqnarray*}
I_{\mu_{1}\ldots\mu_{j}} & = & 0,\;\;\mbox{$j_{2}$ odd}\\
                         & = & I_{j_{1}j_{2}}N(j_{2})T_{\mu_{j_{1}+1}\ldots
                                                \mu_{j}},\;\;
                                    \mbox{$j_{2}$ even},
\end{eqnarray*}
where
\begin{eqnarray*}
   T_{\mu_{j_{1}+1}\ldots\mu_{j}} & = &
                   g^{D-1}_{\mu_{j_{1}+1}\mu_{j_{1}+2}}\ldots
                   g^{D-1}_{\mu_{j-1}\mu_{j}}+
               \mbox{\small all permutations of}\;\{\mu_{j_{1}+1}\ldots
                  \mu_{j}\}\\
 N(j_{2}) & = & g^{\mu_{j_{1}+1}\mu_{j_{1}+2}}\ldots
                  g^{\mu_{j-1}\mu_{j}}T_{\mu_{j_{1}+1}\ldots\mu_{j}}.
\end{eqnarray*}
Here we assumed without loss of generality that the first $j_{1}$
indices lie in the parallel space. We omitted
these indices on the right hand side of the
above equation because they all have fixed value
$\mu_{1}=\ldots=\mu_{j_{1}}=0$
(the parallel space is one-dimensional
in this simple example). So, the expression on the right hand side
is a $j_{2}$-rank tensor in the orthogonal space.
This tensor structure is given by the totally symmetric tensor
$T$, constructed as the symmetric product of metric tensors in
the orthogonal space. $N(j_{2})$ normalizes the tensor $T$.
With $g^{D-1}_{\mu\nu}$ we
denote this metric tensor in the orthogonal space.
One has $g^{D-1\mu}_{\mu}=g^{D-1}_{\mu\nu}g^{\mu\nu}=D-1$.

In the derivation of this formula the celebrated 'existence theorem of
associated functions' has been used \cite{Carl}.

The calculation of higher powers of propagators can now be easily done by
applying appropriate differential operators with respect to masses to the
result expressed in ${\cal R}$-functions \cite{I}.

The next step in our examination of one-loop integrals is the calculation
of the three-point functions. These form factor functions have infrared and
on-shell singularities. They are, therefore, a more sophisticated test of our
method.
Nevertheless
they can be calculated following the same route as
for the 2-point functions \cite{II}. The same is true for arbitrary $n$-
point functions \cite{III,JF,thesis}.

In the following we shortly recapitulate the calculation of the 3-point
functions.
We use the following notation for the one-loop three-point scalar integral
\begin{equation}
I=\int d^{D}l \frac{1}{P_{0}P_{1}P_{2}},
\end{equation}
where
\begin{eqnarray*}
P_{0} & = & l_{0}^{2}-l_{1}^{2}-l_{\bot}^{2}-m_{0}^{2}+i\eta\\
P_{1} & = & (l_{0}+q_{10})^{2}-l_{1}^{2}-l_{\bot}^{2}-m_{1}^{2}+i\eta\\
P_{2} & = & (l_{0}+q_{20})^{2}-(l_{1}+q_{21})^{2}
 -l_{\bot}^{2}-m_{2}^{2}+i\eta,
\end{eqnarray*}
so that $l_{0}$ and $l_{1}$
span the parallel space (the now two-dimensional
 linear span of the
exterior momenta), while $l_{\bot}$, as before, is the modulus of the
loop-momentum in the
orthogonal space.

The resulting non-trivial integrations are
\begin{equation}
I=C\int_{-\infty}^{+\infty}dl_{0}\int_{-\infty}^{+\infty}dl_{1}
\int_{0}^{+\infty}dl_{\bot}\frac{l_{\bot}^{D-3}}{P_{0}P_{1}P_{2}},
\end{equation}
where $C=2\pi^{\frac{D-2}{2}}/\Gamma(\frac{D-2}{2})$.

With the help of a partial fraction
we have three summands to calculate.
The $l_{\bot}$-integration is
easy to do. The next integration can be done via the residue theorem
and the remaining integrals will again be interpreted as
${\cal R}$-functions.

Introducing the abbreviations
{\small
\begin{displaymath}
\begin{array}{lcllcllcl}
a_{01} & = & -2q_{10} & b_{01} & = & 0 & c_{01} & = & m_{1}^{2}-
m_{0}^{2}-q_{10}^{2}\\
a_{02} & = & -2q_{20} & b_{02} & = & 2q_{21} & c_{02} & = & m_{2}^{2}
-m_{0}^{2}-q_{20}^{2}+
q_{21}^{2}\\
a_{11} & = & 2q_{10} & b_{11} & = & 0 & c_{11} & = & -m_{1}^{2}+
m_{0}^{2}-q_{10}^{2}\\
a_{12} & = & 2(q_{10}-q_{20}) & b_{12} & = & 2q_{21} & c_{12} & = & -
m_{1}^{2}+m_{2}^{2}
-(q_{10}+q_{20})^{2}+q_{21}^{2}\\
a_{21} & = & 2q_{20} & b_{21} & = & -2q_{21} & c_{21} & = & -
m_{2}^{2}+m_{0}^{2}-q_{20}^{2}
+q_{21}^{2}\\
a_{22} & = & 2(q_{20}-q_{10}) & b_{22} & = & -2q_{21} & c_{22} & = &
-m_{2}^{2}+m_{1}^{2}
-(q_{10}+q_{20})^{2}+q_{21}^{2},
\end{array}
\end{displaymath}
}
we find
the final result to be
\begin{eqnarray}
I & = &C {\rm B}(2\varepsilon,1)B(1-\varepsilon,\varepsilon)
\sum_{i=0}^{2}\sum_{l=1}^{4} \alpha_{l}
\frac{A_{i,l}^{-\varepsilon}}{C_{i,l}}\label{psum}\\
 & &
{\cal R}_{-2\varepsilon}(\varepsilon,\varepsilon,1;
                         -y_{i,l}^{+},-y_{i,l}^{-},D_{i,l}/C_{i,l})
                 \nonumber\\
\mbox{where} & & \nonumber\\
  y_{i,l}^{+} &  = & -\frac{B_{i,l}}{2A_{i,l}}+\sqrt{\frac{B_{i,
l}^{2}}{4A_{i,l}^{2}}
                                                     -\frac{m_{i}^{2}-
i\eta}{A_{i,l}}}\nonumber\\
  y_{i,l}^{-} &  = & -\frac{B_{i,l}}{2A_{i,l}}-\sqrt{\frac{B_{i,
l}^{2}}{4A_{i,l}^{2}}
                                                     -\frac{m_{i}^{2}-
i\eta}{A_{i,l}}}\nonumber\\
  \alpha_{l} & = & \left\{
                      {+1, l=1,2 \atop -1, l=3,4}
                   \right\} \nonumber\\
  A_{i,1}=A_{i,3} &  = & \frac{a_{i,1}+b_{i,1}}{a_{i,1}-b_{i,1}}\nonumber\\
  A_{i,2}=A_{i,4} &  = & \frac{a_{i,2}+b_{i,2}}{a_{i,2}
                                                       -b_{i,2}}\nonumber\\
  B_{i,1}=-B_{i,3} &  = & \frac{c_{i,1}}{a_{i,1}-b_{i,1}}\nonumber\\
  B_{i,2}=-B_{i,4} &  = & \frac{c_{i,2}}{a_{i,2}
                                                       -b_{i,2}}\nonumber\\
  C_{i,1}=-C_{i,3} &  = &  a_{i,1}(a_{i,2}-b_{i,2})+a_{i,2}(b_{i,1}-a_{i,1})
                          \nonumber\\
  C_{i,2}=-C_{i,4} &  = &  a_{i,2}(a_{i,1}-b_{i,1})+a_{i,1}(b_{i,2}-a_{i,2})
                          \nonumber\\
  D_{i,1}=D_{i,3} &  = &  c_{i,1}(a_{i,2}-b_{i,2})+c_{i,2}(b_{i,1}-a_{i,1})
                          \nonumber\\
  D_{i,2}=D_{i,4} &  = &  c_{i,2}(a_{i,1}-b_{i,1})+c_{i,1}(b_{i,2}-a_{i,2})
  .
                          \nonumber
\end{eqnarray}

We did the calculation of the scalar integral in three steps:
\begin{itemize}
   \item[1.]
      The $l_{\bot}$-integration, followed by a shift in $l_{0},l_{1}$
      to have the $z_{i}$ in a standard form.

   \item[2.]
      A further shift $l_{1}\rightarrow \tilde{l_{1}}=l_{1}+l_{0}$
linearizes the
      $z_{i}$'s and allows for an application of the residue-theorem
      to do the $l_{0}$-integration.

   \item[3.]
      Interpretation of the remainig integration as a representation of
       a ${\cal R}$
      function.
\end{itemize}
Arbitrary tensor integrals involve additional powers $l_{0}^{n_{1}}
l_{1}^{n_{2}}
                                                      l_{\bot}^{2n_{3}}$ in
the numerator so that we have to modify the three steps as follows:
\begin{itemize}
   \item[1.]
      The $l_{\bot}$-integration now gives $-B(1+n_{3}-\varepsilon,
-n_{3}+\varepsilon)
                                            (-z_{i})^{-\varepsilon+n_{3}}$.
      The shift in $l_{0}$ and $l_{1}$ now introduces a (double) binomial sum
of
      powers from $l_{0}^{0}l_{1}^{0}$ to $l_{0}^{n_{1}}l_{1}^{n_{2}}$ in the
      numerator.
   \item[2.]
      The further shift $l_{1}\rightarrow l_{0}+l_{1}$ changes the powers in
the numerator
      to have values from
$l_{0}^{0}\tilde{l_{1}}^{0}$ to $l_{0}^{n_{1}+n_{2}}\tilde{l_{1}}^{n_{1}}$.
      The application of the residue theorem inserts the residue also
      in the numerator, so that the remaining $\tilde{l_{1}}$ integration
involves powers
      $\tilde{l_{1}}^{0}\ldots\tilde{l_{1}}^{n_{1}+2n_{2}}$ in the numerator.
   \item[3.]
      The remaining integrals are of the form $\int_{0}^{\infty}dy
                                                \frac{y^{k}(uy^{2}+vy+
w)^{-\varepsilon+n_{3}}}{ry+s}$
      so that we will end up with results of the form
      $\sim B(k+1,2\varepsilon-2n_{3}-k)\frac{u^{-\varepsilon+n_{3}}}{r}
      $ $ {\cal R}_{-2\varepsilon+2n_{3}+k}
       (\varepsilon-n_{3},\varepsilon-n_{3},1;-y_{+},-y_{-},s/r)$.
\end{itemize}

Applying reduction formulas for ${\cal R}$-functions all tensor integrals
are now expressible in a standard set of well-known functions
\cite{II}. Integrals involving higher powers of propagators can again be
obtained via differentiation formulas.

\subsection{The Singular Case}
In evaluating the three- and higher $n$-point functions one counters more than
just the $UV$-singularities. There
are on-shell and infrared singularities as well. These divergences
always appear as parallel-space divergences. Here they enter the
calculations in the form of endpoint singularities. So they can be
systematically handled by applying {\em Hadamard's finite part\/}
to them \cite{II}.
\section{2-loop Functions}
In this section we start with the calculation of
massive scalar 2-loop functions. We start with the 2-point case as a
more detailed example and present the results for the two different
topologies which appear in the 3-point case.

\subsection{The 2-point Function}
Let us first introduce some notation.
The expression for the normalized two-loop two-point function
is
\begin{displaymath}
   I(q^{2})=-\frac{q^{2}}{\pi^{4}}\int d^{4}l_{1}\int d^{4}l_{2}
   \prod_{i=1}^{5}\frac{1}{P_{i}}.
\end{displaymath}

According to our method
of separating parallel and orthogonal
space variables we rewrite
the integral as
\begin{equation}
   I(q^{2})=-\frac{q^{2}}{\pi^{4}}\int d^{4}l_{1}\int d^{4}l_{2}
   \prod_{i=1}^{5}\frac{1}{{\cal P}_{i}},\label{int}
\end{equation}
where the inverse propagators ${\cal P}_{i}$ are now defined by
     \begin{eqnarray*}
      {\cal P}_{1} & = & l_{10}^{2}-l_{1\bot}^{2}-m_{1}^{2}+i\varrho\\
      {\cal P}_{2} & = & (l_{10}+q_{0})^{2}-l_{1\bot}^{2}-m_{2}^{2}
+i\varrho\\
      {\cal P}_{3} & = & (l_{10}+l_{20})^{2}-l_{1\bot}^{2}-l_{2\bot}^{2}
-2l_{1\bot}l_{2\bot}z-m_{3}^{2}+i\varrho\\
      {\cal P}_{4} & = & (l_{20}-q_{0})^{2}-l_{2\bot}^{2}-m_{4}^{2}
+i\varrho\\
      {\cal P}_{5} & = & l_{20}^{2}-l_{2\bot}^{2}-m_{5}^{2}+i\varrho,
     \end{eqnarray*}
and we choose to be in the rest frame of the exterior momentum
$q\equiv q_{0}$.

The loop momenta $l_{1}$ and $l_{2}$ satisfy
\begin{eqnarray*}
   l_{1} & \equiv & (l_{10},\vec{l}_{1\bot})\\
   l_{2} & \equiv & (l_{20},\vec{l}_{2\bot})\\
   l_{1\bot}      & := & \mid\vec{l}_{1\bot}\mid\\
   l_{2\bot}      & := & \mid\vec{l}_{2\bot}\mid\\
   \vec{l_{1}}\cdot\vec{l_{2}} & = & l_{1\bot}l_{2\bot}z,
\end{eqnarray*}
where the space-like $\vec{l}_{1\bot}$ and $\vec{l}_{2\bot}$
span the orthogonal space.

Note that we have to introduce an angular variable for the relative angle
between $\vec{l}_{1\bot}$ and $\vec{l}_{2\bot}$. We have to do so because
we are not allowed to choose $\vec{l}_{1\bot}\cdot \vec{l}_{2\bot}=0$.
This choice would fail to give the same results for DR and ordinary Riemann
integration in the case of finite integrals.

The integration over the orthogonal space involves six integrations.
Three of them give the product of the volumes of
the unit spheres $S^{1}$ and $S^{2}$ as a prefactor
because only one non-trivial angular integration is
involved in the scalar-product $\vec{l_{1}}\cdot\vec{l_{2}}$.
As a consequence we are left with only five non-trivial integrations,
two ($l_{10}$ and $l_{20}$) from the parallel space and three
($l_{1\bot},l_{2\bot}$ and $z$)
coming from the orthogonal space.
We are left with the the following integrations
to perform
\begin{equation}
   I(q^{2})=\frac{-8q^{2}}{\pi^{2}}\int_{-\infty}^{\infty}dl_{10}
   \int_{-\infty}^{\infty}dl_{20}\int_{0}^{\infty}dl_{1\bot}
   \int_{0}^{\infty}dl_{2\bot}l_{1\bot}^{2}l_{2\bot}^{2}
   \int_{-1}^{1}dz \prod_{i=1}^{5}\frac{1}{{\cal P}_{i}}.\label{int1}
\end{equation}

As usual $\varrho$ is small and strictly positive and the integrations
are well defined for $\varrho\not= 0$. Because we do not Wick rotate
we keep the $i\varrho$-prescription throughout the whole calculation.
At the end of the calculation we will find the limit $\varrho
                                                      \rightarrow 0$
to be well-behaved in every  case.

This `art' of calculating complicated two-loop
functions can be described as transforming
the integrand so that it can be recognized as an integral representation
of special functions \cite{DB}.
Usual methods use Wick rotations, so an
analytical continuation to physical values at the end of the calculation
is always understood. As a consequence, to have sensible results, the
final expressions must be given
in terms of known special functions in order
to manage the analytic continuation back.
The $i\varrho$-prescription is then
re-installed in order to make this continuation unique. In our
approach we avoid altogether the need to Wick rotate.
No analytic continuation at
the end of the calculation is necessary and sensible results can be
obtained by numerical evaluation of the final integral representation.

Let us proceed with the integrations (a detailed discussion can be found
in \cite{2l2p}).
To do so, we first integrate $z$, the cosine of the angle between
$\vec{l_{1}}$ and $\vec{l_{2}}$.

The integral becomes
\begin{eqnarray*}
   I(q^{2}) & = & \frac{-8q^{2}}{\pi^{2}}\int
dl_{10}dl_{20}dl_{1\bot}dl_{2\bot}
                  \frac{l_{1\bot}l_{2\bot}}{2}
    (\frac{1}{{\cal P}_{1}{\cal P}_{2}{\cal P}_{4}{\cal P}_{5}})\\
 & &  \times  \left[
        \log((l_{10}+l_{20})^{2}
   -m_{3}^{2}+i\varrho-(l_{1\bot}+l_{2\bot})^{2})\right. \\
 & & \left.
-\log((l_{10}+l_{20})^{2}-m_{3}^{2}+i\varrho-(l_{1\bot}-l_{2\bot})^{2})
              \right].
\end{eqnarray*}

Expanding the logarithms and using symmetry properties of the integrand
one finds with the help of the residue theorem
the simple result
\begin{eqnarray*}
   I(q^{2}) & = & {\cal C}\int_{-\infty}^{\infty}dl_{10}
\int_{-\infty}^{\infty}dl_{20}
                  \frac{1}{(w_{1}^{2}-w_{2}^{2})(w_{4}^{2}-w_{5}^{2})}\times\\
            & & \times  \left[
                          {\cal L}_{325}-{\cal L}_{315}+{\cal L}_{314}-
{\cal L}_{324}
                                       \right]\\
  {\cal C} &  = & \frac{4q^{2}}{1}\\
   {\cal L}_{ijk} &  = & \log(w_{i}+w_{j}+w_{k})\\
   w_{1}^{2}-w_{2}^{2} &  = & -m_{1}^{2}+m_{2}^{2}-q^{2}-2l_{10}q\\
   w_{4}^{2}-w_{5}^{2} &  = & -m_{4}^{2}+m_{5}^{2}+q^{2}-2l_{20}q,
\end{eqnarray*}
or in dimensionless variables $x=l_{10}/q$, $y=l_{20}/q$
\begin{eqnarray}
   I(q^{2}) & = & {\cal C^{\prime}}\int_{-\infty}^{\infty}dx
   \int_{-\infty}^{\infty}dy
   \frac{1}{(w_{1}^{2}-w_{2}^{2})(w_{4}^{2}-w_{5}^{2})}\times\nonumber\\
     & & \times  \left[
      {\cal L}_{325}-{\cal L}_{315}+{\cal L}_{314}-{\cal L}_{324}
                                       \right]\label{res}\\
  {\cal C^{\prime}} &  = & 4\nonumber\\
   {\cal L}_{ijk} &  = & \log(w_{i}+w_{j}+w_{k})\nonumber\\
   w_{1}^{2}-w_{2}^{2} &  = & -z_{1}^{2}+z_{2}^{2}-1-2x\nonumber\\
   w_{4}^{2}-w_{5}^{2} &  = & -z_{4}^{2}+z_{5}^{2}+1-2y,\nonumber
\end{eqnarray}
where $z_{i}^{2}:=m_{i}^{2}/q^{2}$ replaces $m_{i}^{2}$ in each
$w_{i}$:
\begin{eqnarray*}
   w_{1} &  = &  \sqrt{x^{2}    -z_{1}^{2}+i\varrho}\\
   w_{2} &  = &  \sqrt{(x+1)^{2}-z_{2}^{2}+i\varrho}\\
   w_{3} &  = &  \sqrt{(x+y)^{2}-z_{3}^{2}+i\varrho}\\
   w_{4} &  = &  \sqrt{(y-1)^{2}-z_{4}^{2}+i\varrho}\\
   w_{5} &  = &  \sqrt{y^{2}    -z_{5}^{2}+i\varrho}.
\end{eqnarray*}
Note that for each $w_{i}$ we have $0<\arg(w_{i})<\pi/2$ so that the
limit $\varrho\to 0$ turns out to be well defined.
A detailed discussion of this result can be found in
\cite{2l2p}. This result was tested and confirmed by analytical and
numerical checks and was widely used in comparison with other
numerical routines \cite{A92,jap} and asymptotic expansions \cite{Dav}.
\subsection{The 3-point Functions}
In the following we will
present a integral representation for 2-loop 3-point functions similar to
the one for the 2-loop 2-point function.

\begin{figure}[tbhp]
\begin{picture}(5,5)
   \put(0,2.5){\line(1,0){1}}
   \put(1,2.5){\line(1,1){2.1}}
   \put(1,2.5){\line(1,-1){2.1}}
   \put(2,3.5){\line(1,-3){1}}
   \put(3,4.5){\line(-1,-3){1}}
   \put(1.5,3.5){1}
   \put(1.5,1.5){2}
   \put(2.5,4.5){3}
   \put(2.5,0.5){4}
   \put(3.5,4){5}
   \put(3.5,1){6}
\end{picture}
\caption[The Cross Diagram]{\label{tlc} The Cross Diagram, the labels
denote the propagators}
\end{figure}
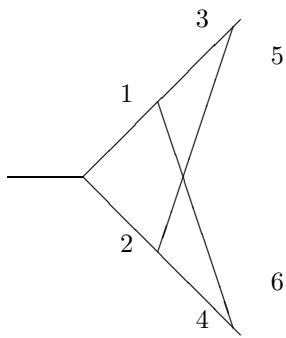

\begin{figure}[tbhp]
\begin{picture}(5,5)
   \put(0,2.5){\line(1,0){1}}
   \put(1,2.5){\line(1,1){2.1}}
   \put(1,2.5){\line(1,-1){2.1}}
   \put(2,3.5){\line(0,-1){2}}
   \put(3,4.5){\line(0,-1){4}}
   \put(1.5,3.5){2}
   \put(1.5,1.5){1}
   \put(2.5,4.5){5}
   \put(2.5,0.5){4}
   \put(2.1,2.5){3}
   \put(3.1,2.5){6}
\end{picture}
\caption[The Planar Diagram]{\label{tlp} The Planar Diagram, the labels
denote the propagators}
\end{figure}
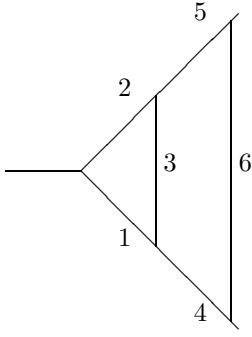

In the 3-point case, we have two scalar master diagrams according to the
two possible different topologies, Fig.(\ref{tlc}) and Fig.(\ref{tlp}).
We will find the amazing result that for {\em both \/} topologies
a threefold integral representation can be given over the same simple
functions as in the 2-point case. Only the number of terms is different
(from 36 for Fig.(\ref{tlp}) to 80 for Fig.(\ref{tlc})).

For 3-point functions additional problems may arise from on-shell
singularities. As we will see our method offers a very systematic handling
for these divergences. Because only endpoint singularities arise in our
calculations, it seems very appropriate to extract these singularities
via principle value prescriptions. A brief introduction to this
method is given in \cite{2l3p}.

It is an amazing consequence of our method that the crossed diagram
Fig.(\ref{tlc})
is of the same order of difficulty as the
parallel diagram Fig.(\ref{tlp}).
This is a consequence of the partial fractionizing, which assigns, roughly
speaking, twice the number of terms to the crossed diagram as to the planar
one.
We begin with a short discussion of the planar function.

The generic function is a function of nine variables (six masses and
three kinematical variables). First we give the explicit expressions
for the propagators:
\begin{eqnarray}
   P_{1} & = & (l_{0}+q_{10})^{2}-l_{1}^{2}-l_{\bot}^{2}-m_{1}^{2}+i
\eta\nonumber\\
   P_{2} & = & (l_{0}-q_{20})^{2}-(l_{1}-q_{21})^{2}-l_{\bot}^{2}-m_{2}^{2}+
i\eta\nonumber\\
   P_{3} & = & (l_{0}+k_{0})^{2}-(l_{1}+k_{1})^{2}-l_{\bot}^{2}-k_{\bot}^{2}
-2l_{\bot}k_{\bot}z-m_{3}^{2}+i\eta\nonumber\\
   P_{4} & = & (k_{0}-q_{10})^{2}-k_{1}^{2}-k_{\bot}^{2}-m_{4}^{2}+
i\eta\nonumber\\
   P_{5} & = & (k_{0}+q_{20})^{2}-(k_{1}+q_{21})^{2}-k_{\bot}^{2}-m_{5}^{2}+
i\eta\nonumber\\
   P_{6} & = & k_{0}^{2}-k_{1}^{2}-k_{\bot}^{2}-m_{6}^{2}+i\eta.\nonumber
\end{eqnarray}
Again $z$ denotes the angle between $l_{\bot}$
and $k_{\bot}$ in the orthogonal space.
The expression for the whole integral is then
\begin{equation}
   I=C\int_{-\infty}^{+\infty}dl_{0}dl_{1}dk_{0}dk_{1}\int_{0}^{\infty}
dl_{\bot}dk_{\bot}\int_{-1}^{+1}dz
   \frac{l_{\bot}k_{\bot}}{\sqrt{1-z^{2}}}\frac{1}{P_{1}P_{2}P_{3}
  P_{4}P_{5}P_{6}}.
\end{equation}
We first integrate $z$
and then make the shifts
$l_{0}\rightarrow l_{0}+l_{1}$ and $k_{0}\rightarrow
                                k_{0}+k_{1}$
which locates both the cuts in  $l_{1}$ and $k_{1}$ of  the square-root in
the result of the $z$-integration
in either the upper or lower halfplanes, depending on the sign of
$(l_{0}+k_{0})$.
Furthermore,  all propagators become linear in $l_{0}$ (resp. $k_{0}$).
So we can
do without any difficulties the $l_{1}$ and $k_{1}$ integrations next
using the residue theorem (details of all these calculations are in \cite{2l3p}
).

The explicit form of the remaining propagators is now
\begin{eqnarray}
   P_{1} & = & (l_{0}+l_{1}+q_{10})^{2}-l_{1}^{2}-l_{\bot}^{2}-m_{1}^{2}
+i\eta\nonumber\\
   P_{2} & = & (l_{0}+l_{1}-q_{20})^{2}-(l_{1}-q_{21})^{2}-l_{\bot}^{2}-
m_{2}^{2}+i\eta\nonumber\\
   P_{4} & = & (k_{0}+k_{1}-q_{10})^{2}-k_{1}^{2}-k_{\bot}^{2}
-m_{4}^{2}+i\eta\nonumber\\
   P_{5} & = & (k_{0}+k_{1}+q_{20})^{2}-(k_{1}+q_{21})^{2}-
k_{\bot}^{2}-m_{5}^{2}+i\eta\nonumber\\
   P_{6} & = & (k_{0}+k_{1})^{2}-k_{1}^{2}-k_{\bot}^{2}-m_{6}^{2}
  +i\eta.\nonumber
\end{eqnarray}
Again we use a partial fraction to simplify the structure of the above product
of propagators. Each of the above
propagators involve either $l$ {\em or\/} $k$ variables,
but not both. So we partial fractionize
the $l$ and $k$ propagators separately:
\begin{eqnarray}
\frac{1}{P_{1}P_{2}P_{4}P_{5}P_{6}} & = &
\frac{1}{P_{1}-P_{2}}\left[\frac{1}{P_{2}}-\frac{1}{P_{1}}\right]\times
 \nonumber\\
 & &
\times\left[ \frac{1}{(P_{4}-P_{5})(P_{4}-P_{6})}\frac{1}{P_{4}}+
\frac{1}{(P_{5}-P_{6})(P_{5}-P_{4})}\frac{1}{P_{5}}
 \right.\nonumber\\
 & & + \left. \frac{1}{(P_{6}-P_{5})(P_{6}-P_{4})}\frac{1}{P_{5}}
 \right].
\end{eqnarray}

This decoupling of $l$ and $k$-variables
reflects the nested loop structure of the planar topology of our graph.
For the crossed ladder graph we will not have this decoupling of variables.

Again all possible differences $P_{i}-P_{j}$ are linear
       monomials in $l_{1}$ or $k_{1}$ with poles on the real axis
(if all $\eta$'s are assumed to be equal).

All of these poles give $i\pi$ contributions to the corresponding integrations.
But there is a further pole coming from the one linearized propagator which
remains in each partial fraction summand. This propagator has a pole
which is not located on the real axis. This pole contributes if it is
in the interior of the contour of integration,
and this in turn  is determined by the sign
of $(l_{0}+k_{0})$. For example, if $(l_{0}+k_{0})>0$, we have to close
the contour in the upper halfplane for both the $l_{1}$
and $k_{1}$-integrations.
Then the pole of $P_{1}$ in $l_{1}$ contributes for $l_{0}<-q_{10}$.
This gives a further splitting in the $l_{0}$-integration.

A list of all the resulting monomials,
sufficient for the construction of the full result,
together with the corresponding {\em REDUCE\/} program, will
be published elsewhere \cite{symb}. It can be also obtained from the author
on request.

Having done the $l_{1}$ and $k_{1}$ integrations, one ends with expressions of
the following form for the $l_{\bot}^{2}$ and $k_{\bot}^{2}$ dependence,
\begin{equation}
   \int_{0}^{\infty}dl_{\bot}^{2}dk_{\bot}^{2}
   \frac{1/\sqrt{\mbox{Poly}(l_{\bot}^{2},k_{\bot}^{2})}}{(a_{l}l_{\bot}^{2}
+b_{l})
                                       (a_{k}k_{\bot}^{2}+b_{k})},
\end{equation}
where $\mbox{Poly}(l_{\bot}^{2},k_{\bot}^{2})$ is a polynomial expression
in $l_{\bot}^{2},k_{\bot}^{2}$ of second degree
\begin{equation}
   \mbox{Poly}(x,y) = a (x-x_{+}(y))(x-x_{-}(y)).
\end{equation}
Let us do the $l_{\bot}^{2}$-integration next.
Expressing the integrand in terms of ${\cal R}$-functions
and applying a
 {\em Landen transformation\/}
\cite{Carl} one can express the result in
terms of logarithms and square roots \cite{2l3p}
\begin{equation}
   {\cal R}_{-1}(1/2,1/2,1;z_{+},z_{-},z_{0})={\cal R}_{-1}(1,1;u_{+},u_{-})=
   \frac{\log(u_{+})-\log(u_{-})}{u_{+}-u_{-}},
\end{equation}
where  $u_{+},u_{-}$ are given by
\begin{eqnarray}
   u_{+} & = &
\sqrt{z_{0}^{2}-z_{0}(z_{+}+z_{-})+z_{+}z_{-}}+z_{0}
\sqrt{z_{+}z_{-}}\nonumber\\
   u_{-} & = &
\sqrt{z_{0}^{2}-z_{0}(z_{+}+z_{-})+z_{+}z_{-}}-z_{0}\sqrt{z_{+}z_{-}}.
 \nonumber\\
\end{eqnarray}
where the $z_{i}$ variables are polynomial expressions in the remaining
integration variables.

In the above mentioned programs the
explicit expressions for $\mbox{Poly}(l_{\bot}^{2},k_{\bot}^{2})$
are generated.
{}From them, one can easily determine the resulting $u_{+}$
and $u_{-}$ expressions. So we end
with a representation as a threefold integral for the generic case of
six different masses. In all cases the $l_{0}$ (resp. $k_{0}$)-axis
is divided in
at most two pieces through the splitting of domains
which means that we have four domains to investigate. We have six summands
from the partial fraction and six (three from
the $k_{1}$ integral times two from the $l_{1}$ integral) contributing
residues in each
domain. This gives 36 terms, each one having its own domain
of contribution.

The crossed function is regarded to be much more difficult because of its
non-planar topology. But we will see in a moment that again, with partial
fractions and shifting of momenta, we can express it in terms of the same
function
as for the planar one. Only the number of terms increases.
The crucial observation is that we can do a partial fraction decomposition
in the two propagators involving both loop momenta. Let us first list the
propagators explicitely:
\begin{eqnarray}
   P_{1} & = & (l_{0}+k_{0}-q_{10})^{2}-(l_{1}+k_{1})^{2}-l_{\bot}^{2}
              -k_{\bot}^{2}-2l_{\bot}k_{\bot}z-m_{1}^{2}+i\eta\nonumber\\
   P_{2} & = & (l_{0}+k_{0}+q_{20})^{2}-(l_{1}+k_{1}+q_{21})^{2}-l_{\bot}^{2}
              -k_{\bot}^{2}-2l_{\bot}k_{\bot}z-m_{2}^{2}+i\eta\nonumber\\
   P_{3} & = & (l_{0}-q_{10})^{2}-l_{1}^{2}-l_{\bot}^{2}-m_{3}^{2}+
i\eta\nonumber\\
   P_{4} & = & (k_{0}+q_{20})^{2}-(k_{1}+q_{21})^{2}-k_{\bot}^{2}-m_{4}^{2}
+i\eta\nonumber\\
   P_{5} & = & l_{0}^{2}-l_{1}^{2}-l_{\bot}^{2}-m_{5}^{2}+i\eta\nonumber\\
   P_{6} & = & k_{0}^{2}-k_{1}^{2}-k_{\bot}^{2}-m_{6}^{2}+i\eta.\nonumber
\end{eqnarray}
Then we have
\begin{eqnarray}
   \frac{1}{P_{1}P_{2}} & = & \frac{1}{P_{1}-P_{2}}\left[
                                                      \frac{1}{P_{1}}-
\frac{1}{P_{2}}
                                                   \right] \nonumber\\
    P_{1}-P_{2} & = &  -2(q_{10}+q_{20})(l_{0}+k_{0})
 +2q_{21}(l_{1}+k_{1})-m_{1}^{2}+m_{2}^{2}.
\end{eqnarray}
In the difference $P_{1}-P_{2}$ the orthogonal space dependence cancels
out completely.

In each summand we can make a shift so as to have the same
structure after the
$z$-integration just as for the planar case. Then we can proceed
exactly as for the planar diagram.
The one crucial difference lies in the extra pole $1/(P_{1}-P_{2})$. This is
a pole
which mixes $l$ and $k$ variables. Nevertheless, the $l_{1},k_{1}$
and $l_{\bot}^{2}$
integrations can be done in a similar manner, only the book keeping of the
contributing expressions and the relevant splittings in the $l_{0}$
and $k_{0}$
become rather tedious. This book keeping is done again with the help of
{\em REDUCE\/} and is included in \cite{symb}.

It is remarkable that we end up
with integral representations of the same kind as in the planar case
while only
the number of terms increases. Explicitely, we have three possible
residues after the $l_{1}$ integration. Inserting the $P_{1}-P_{2}$
pole, we have four poles in the following $k_{1}$ integration. The
other two poles in $l_{1}$ only lead to three poles only by the following
$k_{1}$ integration. So we end with $4+3+3=10$ contributing residues.
For the crossed function, we have $2\times 2\times 2=8$ summands
from the partial fractions (instead of $2 \times 3=6$ in the planar
case). Altogether we have 80 terms which contribute.

Let us mention here that possible on-shell divergences which may appear
will be handled by using Hadamard's finite part. A detailed demonstration
of this method will be presented elsewhere \cite{new}
(see also \cite{thesis,2l3p,II}).

Future work will be devoted to implement this
Hadamard subtraction prescription in
our programs. Note that for the
singularities that have been extracted one fixes one or
two integration variables, so that we have only one- or
twofold integral representations for the coefficient of the singularity, which
can be evaluated at least numerically. A list of the relevant coefficients
in the on-shell case (with the ladder (resp. cross-ladder) propagators
assumed massless)
is already included in \cite{symb}.

\section{Conclusions}
We have presented in examples methods which apply to all loop integrals
which may appear in the Standard Model up to the 2-loop level.
We hope that these methods add to the set of  tools used in multiloop
calculations. Apart from a remarkable systemization of 1-loop functions
we have derived integral representations valid for planar and
non-planar massive
2-loop 2- and 3-point functions.
We calculated these results for the generic case, that is as
functions of nine variables ($q_{0}^{2},q_{1}^{2},q_{0}\cdot q_{1}$, six
masses)
in the three-point case. Accordingly, the results turn out to be complicated.
Nevertheless they are found to have a very systematic structure which
reflects itsels
in a
repeated pattern of typical terms. The extraction of
possible infrared singularities is
reduced
to a book keeping of end-point singularities for both topologies. In future
work, one must investigate how much CPU time is necessary for all
the book keeping. We hope that the above integral representations turn out
to be manageable with respect to these practical considerations.

Also one can try to find easier integral representations if one treats
simpler mass cases. There are enormous possibilities for the application
of these results. Almost every calculation in the electro-weak sector
involves integrals with different masses, but typically
one can restrict ones attention to three different masses. So, if the above
integral representation
allows for an evaluation in a reasonable CPU time, this may open the door to
a new era for calculations devoted to a test of the Standard Model
at two-loop order.\\[0.5cm]
{\large Acknowledgements\\}
Thanks are due to many friends and colleagues, especially to David
Broadhurst, Andrei Davydychev, Junpei Fujimoto San and Vladimir Smirnov for
lots of
stimulating discussions. The author likes to thank the organizing committee
of the Workshop on High Energy Physics and Quantum Field Theory
(Sochi, Russia, Oct.~1992) for the hospitality and friendly atmosphere
during this conference.

This work was supported in part by the Deutsche Forschungsgemeinschaft.
\begin{appendix}
\section{Hadamard's Finite Part}
We give a short summary of the properties of {\em Hadamard's
finite part}. Our presentation
follows  \cite{Zee}. We use this distribution to have a systematic
regularization for on-shell and infrared singularities.

Consider the integral
\begin{equation}
F(a,b,\epsilon)=\int_{a+\epsilon}^{b}
f(t)\phi(t)dt, b>a, \epsilon > 0,\label{b1}
\end{equation}
where $\phi$ belongs to an appropriate set of test-functions.
Let $f(t)$ be locally integrable but singular in the neighborhood
of $t=a$.

Let us assume that it is possible to subtract from $F(a,b,\epsilon)$
a finite linear combination of negative powers of $\epsilon$ and
positive powers of $\log(\epsilon)$ such that the remaining function
of $\epsilon$ allows for a well-defined finite limit $\epsilon\to 0_{+}$.
Then, this limit is called Hadamard's finite part of $f$. It defines
a distribution ${\cal H}(f)$. The functions $f$ which allow for
such a subtraction of the singularities are called pseudofunctions.

To subtract the singularities in $\epsilon$ explicitly, one uses
a Taylor expansion of the test-function $\phi(t)=\phi(a)+
\phi^{(1)}(a)(t-a)+\ldots$ at $a$. For a pseudofunction $f$, only a finite
number of terms is not integrable (in the limit $\epsilon
\to 0_{+}$) when the expansion of $\phi$ is
inserted in Eq.(\ref{b1}). Let us assume that the first $k$ terms in the
Taylor expansion are ill-defined. Then the finite part is given by
\begin{equation}
F_{fin}(a,b,0)=\int_{a}^{b}
f(t)(\phi(t)-\phi(a)-\ldots-\phi^{(k)}(a)(t-a)^{k}/k!)dt,
\end{equation}
while the singular part is a divergent combination of negative powers of
$\epsilon$ and $\log(\epsilon)$ given by the integral of the first
$k$ terms.

The only property of the test-function, that is used in the above
definition of Hadamard's finite part is the fact that it is integrable
in the interval $(a,b)$. Therefore, we can define a regularization scheme
by the use of the above procedure. It is
applicable for every integrand
which belongs to the class of pseudofunctions.
This is sufficient
for our applications because the $i\eta$-prescription guarantees every
Feynman integrand to be a pseudofunction.

The scheme assigns a linear combination of negative powers of $\epsilon$
and positive powers of $\log(\epsilon)$ to the singular part. For the
finite part, its result is Hadamard's finite part.
The generalization to iterated integrals is
straightforward, one mainly has to replace the above Taylor expansion
in one variable by an expansion in several variables.
\end{appendix}

\end{document}